\documentclass[a4paper]{easychair}

\newcommand{\easychair}{\sf{easychair}}

\begin{document}

%
\title{Connecting Gr\"{o}bner bases programs with Coq to do proofs in algebra, geometry and arithmetics}

%
\titlerunning{Connecting Gr\"{o}bner bases programs with Coq}

\volumeinfo
	{G. Sutcliffe, A. Voronkov} 
	{2}                         
	{{\easychair} 1.0, 2008}    
	{1}                         
	{1}                         
	{1}                         

%
\author{Lo\"{\i}c Pottier\\
projet Marelle\\
INRIA Sophia Antipolis\\
\url{loic.pottier@sophia.inria.fr}
}

%
\authorrunning{Pottier}

\maketitle

%
\begin{abstract}
We describe how we connected three programs that compute Gr\"{o}bner bases \cite{buch}  to Coq \cite{coq}, to do automated proofs on algebraic, geometrical and arithmetical expressions. The result is a set of Coq tactics and a certificate mechanism \footnote{downloadable at \url{http://www-sop.inria.fr/marelle/Loic.Pottier/gb-keappa.tgz}}.

The programs are: F4 \cite{F4}, GB \cite{GB}, and gbcoq \cite{gbcoq}. F4 and GB are the fastest (up to our knowledge) available  programs that compute Gr\"{o}bner bases. Gbcoq is slow in general but is proved to be correct (in Coq), and we adapted it to our specific problem to be efficient. The automated proofs concern equalities and non-equalities on polynomials with coefficients and indeterminates in R or Z, and are done by reducing to Gr\"{o}bner computation, via Hilbert's Nullstellensatz. We adapted also the results of \cite{harrison-divisibility}, to allow to prove some theorems about modular arithmetics. The connection between Coq and the programs that compute Gr\"{o}bner bases is done using the "external" tactic of Coq that allows to call arbitrary programs accepting xml inputs and outputs. We also produce certificates in order to make the proof scripts independant from the external programs.
\end{abstract}

\section{Introduction}
\label{sect:introduction}

Proof assistants contain now more and more automatic procedures that generate proofs in specific domains. In the Coq system, several tactics exist, for example the {\tt omega} tactic which proves inequalities between linear expressions with integer variables, the {\tt fourier} tactic which does the same thing with real numbers, the {\tt ring} and {\tt field} tactic, which proves equalities between expressions in a ring or a field, the {\tt sos} tactic which proves some inequalities on real polynomials. 
We describe here a new tactic, called {\tt gb}, which proves (non-)equalities in rings using other (non-)equalities as hypotheses. For example $\forall x y:R, x^2+x y = 0, \; y^2+x y = 0 \Rightarrow x+y=0$, or $\forall x:R, x^2 \not = 1 \Rightarrow x \not = 1$.

This tactic uses external efficient programs that compute Gr\"{o}bner bases, and  their result to produce a proof and a certificate.

We wrote such a tactic several years ago \cite{gbjfla}, but using only the gbcoq program, which were rather slow. So the tactic remained experimental and was not included in the Coq system. There are also similar tactics in other proof systems: in hol-light, John Harrison wrote a program that computes Gr\"{o}bner bases to prove polynomial equalities, specially in arithmetics \cite{harrison-divisibility}. This program was adapted in Isabelle by Amine Chaieb and Makarius Wenzel for the same task \cite{isabelle}. 
We show on examples that our tactic is faster.

This paper is organized as follow. In section 2 we explain the mathematical method we use to reduce the problem to Gr\"{o}bner bases computations.
In section 3 we detail the tactic and the way it builds a proof in Coq. In section 4 we show how we connected Coq to the specialized programs that computes Gr\"{o}bner bases. Section 5 details the complete tactics that proves also non-equalities, and section 6 shows how to produce certificates and then save time in the proof script.  In section 7 we give some examples of utilisations of the tactic in algebra, geometry and arithmetics, with comparisons with hol-light\cite{harrison-holhol} . Section 8 contains the conclusion and perpectives of this work.

\section{Hilbert Nullstellensatz}
\label{sect:typesetting}

Hilbert Nullstellensatz shows how to reduce proofs of equalities on polynomials to algebraic computations (see for example \cite{eisenbud} for the notions introduced in this section).

It is easy to see that if a polynomial  $P$ in $K[X_1,\ldots,X_n]$ verifies $P^r = \sum_{i=1}^{s} Q_i P_i$, with $r$ a positive integer, $Q_i$ and $P_i$ also in $K[X_1,\ldots,X_n]$, then $P$ is zero whenever polynomials $P_1,...,P_s$  are zero.

Then we can reduce the proof of $P_1=0,\ldots,P_s=0 \Rightarrow P=0$ to find $Q_1,\ldots,Q_s$ and $r$ such that $P^r = \sum_{i} Q_i P_i$. 

The converse is also true when K is algebraically closed: this is the Hilbert Nullstellensatz. In this case, the method is complete.

Finding $P^r = \sum_{i} Q_i P_i$ can be done using Gr\"{o}bner bases, as we will explain now.

Recall that an {\em ideal} ${\cal I}$ of a ring is an additive sub-group  of the ring such that $a x \in {\cal I}$ whenever $a \in {\cal I}$. 
The ideal {\em generated} by a family of polynomials is the set of all linear combinations of these polynomials (with polynomial coefficients).

A {\em Gr\"{o}bner basis} of an ideal is a set of polynomials of the ideal such that their head monomials (relative to a choosen order on monomials, e.g. lexicographic order, or degree order) generates the ideal of head monomials of all polynomials in the ideal. The main property of a  Gr\"{o}bner basis is that it provides a test for the membership to the ideal: a polynomial is in the ideal iff its euclidian {\em division} by the polynomials of the basis gives a zero remainder. The division process is a generalisation of the division of polynomials in one variable: to divide a polynomial $P$ by a polynomial $a X^\alpha -Q$ we write $P = a X^\alpha S + T$ where $T$ contains no monomial that is multiple of $X^\alpha$. Then change $P$ with $Q S + T$ and repeat divison. The last non zero $T$ is the remainder of the division. To divide a polynomial by a family of polynomials, we repeat this process with each polynomial of the family. In general, the remainder depends on the order we use the polynomials of the family. But with a Gr\"{o}bner basis, this remainder is unique (this is a characteristic property of Gr\"{o}bner basis).

\subsection{Method 1: how to find $Q_1,\ldots,Q_s$ such that $1 = \sum_{i} Q_i P_i$}

Compute a Gr\"{o}bner base of the polynomials $\{t P_i - e_i,\; e_i e_j,\; e_i t\}_{i,j} $ (where $t,e_1,\ldots,e_s$ are new variables) with an order such that $t > X_i > e_i$.

Suppose that, in this basis, there is a polynomial of the form $t-\sum_{i} Q_i e_i$. This polynomial is then in the ideal generated by $\{t P_i - e_i,\; e_i e_j,\; e_i t\}_{i,j} $, so is a linear combination of these polynomials:
 
$t-\sum_{i} Q_i e_i = \sum_{i} h_i (t P_i - e_i)\;+\;\sum_{i j} g_{i j} e_i e_j \;+\; \sum_{i} k_i e_i t$

$e_i$ are formal variables, so we can substitute formally $e_i$ with $t P_i$, and we obtain $t (1-\sum_{i} Q_i P_i) = 0\;+ t^2 (\;\sum_{i j} g_{i j} P_i P_j \;+\; \sum_{i} k_i P_i)$.

Then the coefficient of $t$ in this equation must be zero:
$1-\sum_{i} Q_i P_i = 0$, and we are done.

Note that the polynomials $\{ e_i t,\; e_i e_j\}$ are not necessary, but their presence much speed up the computation of the Gr\"{o}bner basis\footnote{thanks to Bernard Mourrain for this trick}.

\subsection{Method 2: how to find $Q_1,\ldots,Q_s$ and $r$ such that $P^r = \sum_{i} Q_i P_i$ }
Use the standard trick: search to write $1 = \sum_{i} h_i P_i \; + \; h (1-z P)\; (*)$, where $z$ is a new variable. This can be done with the previous method. Suppose we succeed. Let $r$ be the max degree in $z$ of polynomials $h_i$. 

Substitute formally $z$ with $1/P$, and multiply the equation (*) by $P^r$. Then we obtain
$P^r = \sum_{i} Q_i P_i$, as required, where $Q_i = P^r h_i[z \leftarrow 1/P]$

\subsection{Completness}
It is easy to see that methods 1 and 2 are complete in the sense that if $P^r = \sum_{i} Q_i P_i$ holds, there will find such an equation: 
\begin{itemize}
\item method 1: suppose $1 -\sum_{i} Q_i P_i = 0$. Then $t = \sum_{i} Q_i t P_i$, and  $t - \sum_{i} Q_i e_i = \sum_{i} Q_i (t P_i - e_i)$. Hence $t - \sum_{i} Q_i e_i $ belongs to the ideal of which we have computed a Gr\"{o}bner basis. Because of the order we have choosen on variables, this implies that there is a polynomial $t - \sum_{i} h_i e_i $ in the Gr\"{o}bner basis.
\item method 2: suppose $P^r = \sum_{i} Q_i P_i$. We have $1-z^r P^r = (1+z P + \ldots + z^{r-1} P^{r-1}) (1-z P)$. Replacing $P^r$ with $\sum_{i} Q_i P_i$ we obtain $1 = z^r (\sum_{i} Q_i P_i) + (1+z P + \ldots + z^{r-1} P^{r-1}) (1-z P)$.
\end{itemize}

\subsection{Example}

Take $p= x+y,\; p_1 = x^2 +x y,\; p_2 = y^2+x y$. With the previous method, the Gr\"{o}bner basis is:

\[\begin{array}{l}
 t -z y e_0 -z x e_0 -z^2 e_1 -z^2 e_2 -e_0\\
y^2 e_0 -x^2 e_0+z y e_1 -z x e_2 -e_1+e_2\\
y x e_0+x^2 e_0+z y e_2+z x e_2 -e_2\\
e_0^2\\
x e_1 -y e_2\\
e_0 e_1, e_1^2, e_0 e_2, e_1 e_2, e_2^2\\
\end{array}\]

 we obtain $r=2,\; Q_1=1,\; Q_2 = 1$, and then $(x+y)^2= 1 \times (x^2+x y) + 1 \times (y^2 + x y)$. Which proves that $x^2+x y = 0, \; y^2+x y = 0 \Rightarrow x+y=0$.

\section{Proof in Coq }

Coq \cite{coq} is a proof assistant based on type theory, where we can interactively build proofs of {\em goals}, which are logical assertions of the form $\forall H_1:T_1, \ldots, \forall H_n:T_n, C(H1,\ldots,H_n)$. Using tactics, we can simplify the goal, while the system builds the corresponding piece of proof.

Typically we will treate goals of the form:

\begin{verbatim}
  x : Z
  y : Z
  H : x ^ 2 + x * y = 0
  H0 : y ^ 2 + x * y = 0
  ============================
   x + y = 0
\end{verbatim}

Here hypotheses are variables belonging in a ring or a field, and equalities between polynomials.

We explain now how to compute and use the Nullstellensatz equation to build a proof of this goal in Coq. The steps are: syntaxification, Gr\"{o}bner basis computation, and building the proof from the Nullstellensatz equation.

\subsection{Syntaxification}
We begin by building polynomials from the three equations in this goal. This is done in the tactic language of Coq (LTAC, which is a meta-language for computing tactics and executing them) by first computing the list of variables:

\begin{verbatim}
 lv = (cons y (cons x nil))
\end{verbatim}

and the list of polynomials:
\begin{verbatim}

 lp = (cons (Add (Pow (Var 2) 2) (Mul (Var 2) (Var 1)))
      (cons (Add (Pow (Var 1) 2) (Mul (Var 2) (Var 1)))
      (cons (Sub (Add (Var 2) (Var 1)) (Const 0 1)) 
       nil)))
\end{verbatim}

Variables are represented by their rank in the list of variables.
Polynomials are elements of an inductive type, and we can recover the equations by interpreting them in Z with the list of variables. For example, 
\begin{verbatim}
 (interpret (Add (Pow (Var 2) 2) (Mul (Var 2) (Var 1)))
            lv)
\end{verbatim}
evaluates in {\tt x\^{ }2 + x * y}.

We used parts of the code of the {\tt sos}\cite{harrison-sos} tactic, written by Laurent Th\'ery.

\subsection{Calling Gr\"{o}bner basis computation}
We call the external program gb (see section 4) with the list of polynomials; here we choose the program F4 to compute Gr\"{o}bner basis:
\begin{verbatim}
 external "./gb" "jcf2" lp
\end{verbatim}

The result is the term:

\begin{verbatim}
 (cons
    (Pow
      (Add
         (Add Zero
            (Mul
               (Add (Add Zero (Mul (Const 0 1) (Const 1 1)))
                  (Mul (Const 1 1) (Pow (Var 1) 1))) (Const 1 1)))
         (Mul (Const 1 1) (Pow (Var 2) 1))) 
       2)
 (cons (Const 1 1) (cons(Const 1 1) (cons (Const 1 1) nil))))
\end{verbatim}

which has the structure 
\begin{verbatim}
  (cons (Pow p d) (cons c lq))
\end{verbatim}

such that the Nullstellensatz equation holds: \[c p^d = \sum_{q_i \in lq} q_i p_i\]

Here, we have $lq=q_1, q_2$, $q_1=q_2=1$

\subsection{Building the proof from the Nullstellensatz equation}
  
After interpreting the polynomials $q_1$ and $q_2$ in Z using the original list of variables, we get and prove easily the goal

{\tt 
 1 * (x + y)\^{ }2 = 1 * (x \^{ } 2 + x * y) + 1 * (y \^{ } 2 + x * y)
}

by the ring tactic.

To prove the original goal, it is now sufficient to rewrite {\tt x\^{ }2 + x * y} and {\tt y\^{ }2 + x * y} by 0, getting {\tt  1 * (x + y)\^{ }2 = 0}, and, using a simple lemma, we get {\tt x + y = 0} and we are done.

\section{Connecting F4, GB, and gbcoq to Coq }
Coq allows to call arbitrary external programs via a function called "external". It sends Coq terms in xml format (i.e. as tree) to the standard output of the external program, and gets its standard output (also in xml format) as a resulting Coq term. We use this function to compute a Gr\"{o}bner basis of a list of polynomials, via a single interface to three specialized programs: F4, GB, and gbcoq. 
This interface, called "gb" is written in ocaml. It translates the list of polynomials given as standard input in xml format in the format of the choosen program (F4, GB or gbcoq), call it with the good arguments, get its result (a Gr\"{o}bner basis, if no error occured), selects its useful information, translates it in xml and sends it as result to standard output.
More precisely:
\begin{itemize}
\item F4 is a C library, and has only an interface for Maple. We wrote a simple parser of polynomials to use it on command line, helped by J.C. Faug\`{e}re. 
\item GB is also written in C and has a command line interface, or accept inputs in a file; with a Maple-like syntax for polynomials. 
\item Gbcoq is written in ocaml, so is integrated to gb. This program uses an Buchberger-like algorithm which has been extracted from Coq. So it is proven to be correct. We added recently an optimisation which reduces drastically the time to compute Nullstellensatz equations: each time we add a new polynomial during the completion via the reduction of critical pairs, we divide the polynomial that we want to test if it is in the ideal, by the current family of polynomials. If this gives zero, then we stop:, and return the Nullstellensatz coefficients, deduced from the divisions we made. More we also try its powers (up to a parametrized limit). Then, when we have computed the whole Gr\"{o}bner basis, we can compute the Nullstellensatz coefficients, without having to verify that the remaining critical pairs reduce to zero. More, this is often the case that the polynomial reduces to zero with a partial Gr\"{o}bner basis! The time is sometimes divided by 1000 with such a technique, and always much reduced.
Note that such an improvement cannot be made in a blackbox program such as the programs of JC Faug\`ere, which are free but not opensource.

\end{itemize}

\section{The gbR and gbZ tactics in Coq }

We wrote two tactics: gbR for real numbers, gbZ for integers.  The set of integer is not a field, but we can simulate computations in the field of rational numbers using only integers. In this case, the Nullstellensatz equation become $c p^d = \sum_{i} q_i p_i$, where $c$ is an integer, and the $q_i$ have integer coefficients.

We can allow negations of equations in the conclusion. For example $x y = 1 \Rightarrow x \not = 0$. The trick is to replace $x \not = 0$ with $x=0 \Rightarrow 1=0$, which is equivalent to add a new equation in hypotheses, and replace the equation to prove with $1=0$.

In the case of real numbers, we can allow also negations of equations in hypotheses. For example $x^2\not = 1 \Rightarrow x \not = 1 $. This can be done by introducing new variables, remarking that $p \not = 0 \Leftrightarrow \exists t, p*t=1$. In the example, this gives $t(x^2-1)=1 \Rightarrow x \not = 1$. The negation in conclusion can be removed and leads to $t(x^2-1)=1, \; x-1=0 \Rightarrow 1=0$, which is proven using the Nullstellensatz equation $1 = 1 \times (t(x^2-1)-1) + (t+t x)\times(x-1)$

Finally, the tactics use first the program F4. If it fails (for memory limits), then the tactics try GB. If it fails too, then the tactics uses gbcoq. We have also specialised tactics, allowing the user to choose which program to use, between F4, GB, and gbcoq. Indeed, experiments show that no one is better than others.

\section{Certificates}

Once the Nullstellensatz equation is computed, we can change the proof script, replacing the tactic gb with a similar tactic, called "check\_gb" which will not call external programs, but instead it will take as arguments all the components of the Nullstellensatz equation. So, next time we will execute the proof script, for compilation for example, it will not need external Gr\"{o}bner computation\footnote{thanks to Julien Narboux for this suggestion}.
Let us give an example.
Suppose we want to prove:

\begin{verbatim}
 Goal forall x y z:R, x^2+x*y=0 -> y^2+x*y=0 -> x+y=0.
\end{verbatim}

we execute the tactic gbR, which proves the goal, and prints these lines in the standard output of Coq:
\begin{verbatim}
 (* with JC.Faugere algorithm F4 *)
 gbR_begin; check_gbR
 (x + y - 0)
 (List.cons (x * (x * 1) + x * y) (List.cons (y * (y * 1) + x * y) List.nil))
 (List.cons y (List.cons x List.nil))

(lceq
   (Pow
      (Add
         (Add Zero
            (Mul
               (Add (Add Zero (Mul (Const 0 1) (Const 1 1)))
                  (Mul (Const 1 1) (Pow (Var 1) 1))) (Const 1 1)))
         (Mul (Const 1 1) (Pow (Var 2) 1))) 2)
   (lceq (Const 1 1) (lceq (Const 1 1) (lceq (Const 1 1) lnil))))
 .
\end{verbatim}

Then, we can replace the line calling gbR with these tactics lines, which contains no more than the components of the needed Nullstellensatz equation $(x+y)^2= 1 \times (x^2+x y) + 1 \times (y^2 + x y)$, and then need much less time to evaluate, because it doesn't need Gr\"{o}bner basis computation.

\section{Examples}

In this section we give several examples of use of the tactics gbR and gbR.

\subsection{Algebra}

The following examples uses the symetric expressions of coefficients  with roots of a polynomial.

First in degree 3: if $x,y,z$ are the three complex roots of $X^3+a*X^2+b*X+c$ then we have $a=-(x+y+z)$, $b=x*y+y*z+z*x$, and $c=-x*y*z$. And then we can prove that $x+y+z=0 \Rightarrow x*y+y*z+z*x=0 \Rightarrow x*y*z=0 \Rightarrow x=0$, because then the polynomial becomes $X^3$, and has only $0$ as a root.

\begin{verbatim}
Require gbZ.

Goal forall x y z:Z,
  x+y+z=0 -> x*y+y*z+z*x=0 -> x*y*z=0 -> x=0.
gbZ.
Qed.
\end{verbatim}

More complicated, the same thing in degrees 4 and 5:

\begin{verbatim}
Goal forall x y z u:Z,
  x+y+z+u=0 -> 
  x*y+y*z+z*u+u*x+x*z+u*y=0 ->
  x*y*z+y*z*u+z*u*x+u*x*y=0 -> 
  x*y*z*u=0 -> x=0.
gbZ.
Qed.

Goal forall x y z u v:Z,
  x+y+z+u+v=0 -> 
  x*y+x*z+x*u+x*v+y*z+y*u+y*v+z*u+z*v+u*v=0->
  x*y*z+x*y*u+x*y*v+x*z*u+x*z*v+x*u*v+y*z*u+y*z*v+y*u*v+z*u*v=0->
  x*y*z*u+y*z*u*v+z*u*v*x+u*v*x*y+v*x*y*z=0 ->
  x*y*z*u*v=0 -> x^5=0.
gbZ.
Qed.
\end{verbatim}
Last example takes less than 1s with F4 and GB, and gbcoq.
With hol-light, it takes 1s.

\subsection{Geometry}

Desargues theorem is too complicated to be proved with Gr\"{o}bner bases. But Pappus theorem can. We formalize in Coq the set of points in the real plane:

\begin{verbatim}
Open Scope R_scope.
Record point:Type:={
 X:R;
 Y:R}.
\end{verbatim}

Then we give two definitions of colinearity of three points (the theorem is false if we use only the second definition, because of degenerated configurations):
\begin{verbatim}

Definition colinear(C A B:point):=
  exists a:R, 
 (X C)=a*(X A)+(1-a)*(X B) /\ (Y C)=a*(Y A)+(1-a)*(Y B).

Definition colinear2(A B C:point):=
  (X A)*(Y B)+(X B)*(Y C)+(X C)*(Y A)
  =(Y B)*(X C)+(Y C)*(X A)+(Y A)*(X B).

\end{verbatim}

Then we state and prove the Pappus theorem, in a specialized (but without lost of generality) configuration:

\begin{verbatim}
Lemma pappus: forall A B C A' B' C' D E F:point,
  (X A')=0 -> (X B')=0-> (X C')=0 ->
  (Y A)=0 -> (Y B)=0 -> (Y C) = 0 ->
  colinear D A B' -> colinear D A' B ->
  colinear E A C' -> colinear E A' C -> 
  colinear F B C' -> colinear F B' C ->
  colinear2 D E F.
...
gbR_choice 2.
Qed.

\end{verbatim}

In this example, F4 fails, GB takes 9s, and gbcoq takes 3s.
We also tried hol-light with this example, which takes 77s:
\begin{verbatim}
./hol

prioritize_int();;

let t1 = Unix.time();;

int_ideal_cofactors
[`XD -( x4 * XA )`;
`YD -((&1 - x4) * YB1)`;
`XD -( (&1 - x3) * XB)`;
`YD - (x3 * YA1)`;
` XE - x2 * XA`; 
`YE -  (&1 - x2) * YC1`;
` XE - (&1 - x1) * XC`;
` YE - x1 * YA1`; 
` XF - x0 * XB`;
` YF -  (&1 - x0) * YC1`;
` XF - (&1 - x) * XC`;
` YF - x * YB1`]
` XD * YE + XE * YF + XF * YD -(YE * XF + YF * XD + YD * XE)`;;

Unix.time()-.t1;;

\end{verbatim}

The general case of Pappus theorem is too complicated to compute.

\subsection{Arithmetics}
Following the idea of \cite{harrison-divisibility}, we can prove statements about coprimality, gcd and divisions. We have to do some work for that, because the tactic gbZ is not sufficient. But the problem is again an ideal membership one, then solvable by Gr\"{o}bner basis computation. We have written a tactic doing that, called {\tt gbarith}. Here are examples of its use in Coq:

\begin{verbatim}
Definition divides(a b:Z):= exists c:Z, b=c*a.
Definition modulo(a b p:Z):= exists k:Z, a - b = k*p.
Definition ideal(x a b:Z):= exists u:Z, exists v:Z, x = u*a+v*b.
Definition gcd(g a b:Z):= divides g a /\ divides g b /\ ideal g a b.
Definition coprime(a b:Z):= exists u:Z, exists v:Z, 1 = u*a+v*b.

Goal forall a b c:Z, divides a (b*c) -> coprime a b -> divides a c.
gbarith.
Qed.

Goal forall m n r:Z, divides m r -> divides n r -> coprime m n -> divides (m*n) r.
gbarith.
Qed.

Goal forall x y a n:Z, modulo (x^2) a n -> modulo (y^2) a n -> divides n ((x+y)*(x-y)).
gbarith.
Qed.
\end{verbatim}

\subsection{Computation times, comparison with hol-light}

Previous examples, and more we made, show that no one among F4, GB, gbcoq and is better than others. hol-light is sometimes better than F4 and GB, but gbcoq is much better than hol-light. The reason is simple: we often stop computations before obtaining a Gr\"{o}bner basis.

\section{Conclusion}

The ''external'' tactic of Coq is a very good tool to use efficient programs to produce proofs in specific domains. We have shown how to use efficient Gr\"{o}bner bases computations in this context.
The use of certificates should be developped to reduce time of re-verification of proofs. The certificate can be written explicitely in the proof script, as we have shown here, but it could be stored in a cache.
We have shown the interest of using external programs, but also their limits, as soon as it is impossible or difficult to adapt them to specific use of proof systems.
We plan to investigate other decisions procedures, for example polynomial system solving, to produce new tactics in the same spirit.

{\bf Acknowledgements:} we thank anonymous referees for their suggestions on the redaction of this paper and bibliographical completions.
%
\label{sect:bib}
\bibliographystyle{plain}
\bibliography{easychair}

\end{document}